# Static and Dynamic Magnetic Response of Fragmented Haldane-like Spin Chains in Layered $Li_3Cu_2SbO_6$

Changhyun Koo[1], Elena A. Zvereva[3], Igor L. Shukaev[4], Michael Richter[1], Mikhail I. Stratan[3], Alexander N. Vasiliev[3,5,6], Vladimir B. Nalbandyan[4], Rüdiger Klingeler[1,2,*]

[1]*Kirchhoff Institute of Physics, Heidelberg University, INF 227, 69120 Heidelberg, Germany*

[2]*Centre for Advanced Materials (CAM), Heidelberg University, INF 225, 69120 Heidelberg, Germany*

[3]*Faculty of Physics, Moscow State University, Moscow, 119991 Russia*

[4]*Chemistry Faculty, Southern Federal University, Rostov-na-Donu, 344090 Russia*

[5]*Ural Federal University, Ekaterinburg 620002, Russia*

[6]*National University of Science and Technology "MISiS", Moscow 119049, Russia*

**Abstract**

The structure and the magnetic properties of layered $Li_3Cu_2SbO_6$ are investigated by powder X-ray diffraction, static susceptibility, and electron spin resonance studies up to 330 GHz. The XRD data experimentally verify the space group C2/m with halved unit cell volume in contrast to previously reported C2/c. In addition, the data show significant Li/Cu-intersite exchange. Static magnetic susceptibility and ESR measurements show two magnetic contributions, i.e. quasi-free spins at low-temperature and a spin-gapped magnetic subsystem, with about half of the spins being associated to each subsystem. The data suggest ferromagnetic-antiferromagnetic alternating chains with $J_{FM}$ = -285 K and $J_{AFM}$ = 160 K with a significant amount of Li-defects in the chains. The results are discussed in the scenario of fragmented 1D S = 1 AFM chains with a rather high defect concentration of about 17% and associated S = ½ edge states of the resulting finite Haldane chains.

[*] klingeler@kip.uni-heidelberg.de

## 1. Introduction

Unusual ground states and exotic low-energy excitations appearing in low-dimensional spin systems enable experimentally investigating fundamental properties of quantum many particle systems. Uniform one-dimensional (1D) $S = ½$ Heisenberg spin chains with isotropic and predominant nearest-neighbor exchange interaction exhibit a degenerated singlet ground state where spin correlations decay slowly by a power law. [1] The spectrum of triplet excitations is gapless. Even small perturbations such as anisotropy, interchain coupling or alternating exchange couplings can however drastically change the excitation spectrum and result in long-range order or spin-gapped ground states. For integer spin systems, e.g. $S = 1$ chains, the singlet ground state exhibits exponentially decaying spin correlations and there is a robust energy gap in the spin excitation spectrum. [2] $S = 1$ spin chains are well described in terms of the valence-bond-solid model where spins at each site are decomposed into two $S = ½$ which couple antiferromagnetically to corresponding $S = ½$ spins at the two neighboring chain sites. [3]

Non-magnetic localized defects can play a decisive role in low-dimensional quantum spin systems as they may strongly affect the ground states and the excitation spectra. A prominent example are the $CuO_2$ layers of high-temperature cuprate superconductors where Zn-doping yields localized magnetic impurities in the primarily undoped two-dimensional Heisenberg $S = ½$ antiferromagnet. [4] An even more spectacular effect is seen in the hole-doped layers of $La_{1-x}Sr_xCuO_4$ with $x > 0$, where the interplay of charge and spin degrees of freedom results in an increase of the magnetic ordering temperature when diluting the spin system by introducing non-magnetic Zn defects. [5] Zn-defects have also been applied to control the mean free path of magnetic excitations in $La_2Cu_{1-z}Zn_zO_4$ [6]. In the frustrated $S = ½$ quantum spin chain Heisenberg magnet $CaCu_{2-x}Zn_xO_3$, defect-induced spin states stabilize the long-range magnetically ordered ground state. [7,8] Theoretically, enhancement of local antiferromagnetic correlations in low-dimensional Heisenberg antiferromagnets is described in terms of the 'pruned' resonating-valence-bond (RVB) picture.[9,10]. In 1D spin systems non-magnetic impurities may effectively truncate magnetic correlations by decoupling 1D chains into finite fragments. [11] Experimental studies on finite quantum spin chain systems have allowed, e.g., investigating the spin correlation length in the Haldane chains in $Y_2BaNi_{0.96}Mg_{0.04}O_5$ [12,13], magnetic order in $Li_2Cu_{1-z}Zn_zO_2$ [14], $CuO_2$ $S = ½$ spin chains under variation of the number of charged defects or with quenched defect disorder [15,16].

Even ferromagnetism can evolve by Li- and K-intercalation in a $S = ½$ spin gap system as has been observed in Vanadium oxide nanotubes [17,18].

In the recent years, several realizations of a non-magnetic ground state with a spin-gap behavior have been found in honeycomb lattice systems such as $Na_2Cu_2TeO_6$ [19,20], $Na_3Cu_2SbO_6$ [21,22,23], and $Cu_5SbO_6$ ($Cu^+_3Cu^{2+}_2Sb^{5+}O_6$) [24]. In these systems, the interplay of particular orbital arrangement, lattice distortion and frustration causes quasi-one-dimensional (1D) structures or weakly interacting magnetic dimers. In contrast to well-known superexchange between edge-sharing $CuO_6$ octahedra in honeycomb layers, in these examples the dominating exchange pathways involve a Cu-O-O-Cu superexchange mechanism where $Cu^{2+}$-ions are separated by non-magnetic $Sb^{5+}$ ions. In the present work, we show that in the related system $Li_3Cu_2SbO_6$ quasi-1D ferromagnetic-antiferromagnetic $S = ½$ chains evolve which resemble fragmented $S = 1$ Haldane chains. In addition, we report experimental verification of the reindexed structure in terms of the space group C2/m with halved unit cell volume [25] as compared to C2/c reported by West el al. who firstly prepared $Li_3Cu_2SbO_6$ [26].

## 2. Experimental

Two samples of nominally identical composition $Li_3Cu_2SbO_6$ were prepared by different routes using reagent-grade starting materials. They were characterized by powder X-ray diffractions (XRD) with Cu $K_\alpha$-radiation using an ARL X'tra diffractometer equipped with a solid-state Si(Li) detector. The crystal structures were refined by the Rietveld method using GSAS + EXPGUI suit [27,28].

The sample labeled SSS (solid-state synthesis) is a reproduction of the preceding solid-state preparation [25] at final temperature of 1000 °C (1025 °C in Ref. 25). The sample labeled IEP (ion-exchange preparation) was obtained from $Na_3Cu_2SbO_6$ [29] by treatment in excess molten $LiNO_3$ for 2 h at 290 °C followed by washing with methanol and drying in a desiccator. Lattice parameters of both samples are very similar to each other and to those reported previously (Table 1). However, the XRD pattern of the IEP sample, in contrast to the SSS sample and to the sodium precursor, is diffuse, obviously due to stacking faults and/or cracks developed during the ion exchange (Fig. 1).

High-temperature preparation of $Li_3Cu_2SbO_6$ resulted in partial Cu/Li inversion: a considerable part of copper is found on alkali sites and *vice versa* [25]. This is not the case with $Na_3Cu_2SbO_6$ because of the much larger size difference between $Na^+$ and $Cu^{2+}$, and its structure is completely ordered. [29] Since the rigid part of structure usually remains intact during the low-temperature ion exchange, we expected the IEP sample to be completely ordered, too, in contrast to the SSS sample. However, the Rietveld refinement results in an essentially identical degree of inversion (Table 1). Details of the refinement and the structural results are listed in Tables 2, S1, and S2 whereas experimental and calculated diffraction patterns are compared in Fig. 1.

Since preliminary measurements showed that the properties of the two samples are essentially identical, the SSS sample was selected for detailed studies of its static and dynamic magnetic properties. The static magnetic susceptibility $\chi$ = M/B has been investigated in the temperature range 2 K – 300 K at 0.1 T with a SQUID magnetometer (MPMS XL-5, Quantum Design). X-band electron spin resonance (X-ESR) spectra have been obtained by means of a CMS 8400 (ADANI) spectrometer at fixed frequency of about 9.4 GHz and in magnetic fields $0 \leq B \leq 0.7$ T. Measurements have been performed on powder samples at T = 7 - 300 K. High-frequency electron spin resonance (HF-ESR) measurements were performed on fixed powder in a wide frequency range from 80 GHz to 330 GHz, and in a temperature range from 3 K to 270 K. The powder sample mixed with eicosane was placed into a holder ring and wrapped by kapton tape. A millimeterwave vector network analyzer (AB Millimetre) was used as a source and detector of stable microwave, and a superconducting magnet provides magnetic field up to 18 T [30]. HF-ESR powder spectra were simulated with the SIM software by H. Weihe [31].

## 3. Magnetization and ESR studies

The static magnetic susceptibility $\chi^{exp}$ = M/B of $Li_3Cu_2SbO_6$ shows two different magnetic contributions as particularly visible in $\chi^{-1}$ (see Fig. 2). To be specific, the susceptibility obeys Curie-Weiss-like behavior, $\chi_{CW}$ = C/(T+Θ) + $\chi_0$, in two different temperature regimes: at high temperatures T > 200 K (regime II) and at low temperatures T < 30 K (regime I), respectively, with Curie constants $C^I$ and $C^{II}$, Weiss temperatures $Θ^I$ and $Θ^{II}$, and a temperature independent contribution $\chi_0$. The different regions of Curie-Weiss-like behavior are particularly evident when $1/\chi^{exp}$ is considered which displays nearly perfect linearity both at high and low temperatures, respectively, and a crossover region in-between. In the Curie-Weiss approximation, such bending is associated with an increase of the effective magnetic moment. In order to illustrate the magnetic response in more detail, we have approximated the low-temperature susceptibility by means of the above-mentioned modified Curie-Weiss law (see the dashed lines in Fig. 2). The analysis yields $\Theta^I \sim 0$ K and implies that the number of $Cu^{2+}$ spins S = 1/2 associated with $C^I$ is about 0.8/f.u., i.e. more than 40% of spins can be considered quasi-free obeying a Curie-like behavior ($\Theta \sim 0$ K). It is straightforward to assume that $Cu^{2+}$-ions at Li-sites are associated to quasi-free magnetic moments. Our XRD analysis (cf. table S1) implies 0.34 $Cu^{2+}$-ions at Li1- and Li2-sites which yields 0.34 quasi-free Cu-spins S = ½ per f.u.. Even if a few additional magnetic defects not resulting from intersite exchange are considered, this number clearly falls below the number of quasi-free spins found in the experimental data.

The difference $\chi^{exp} - \chi_{CW}^I$ shown in the inset of Fig. 2 displays thermally activated behavior, i.e. a spin-gapped contribution $\chi_\Delta$ to the magnetic susceptibility. The maximum of $\chi_\Delta$ appears at T ≈ 80 K. The kink in $1/\chi^{exp}$ signals the thermally activated crossover of the remaining spins (i.e. those which cannot be considered quasi-free) from a non-magnetic to a magnetic state.

X-ESR data corroborate well with the picture yielded from the static magnetic susceptibility revealing the presence of two different magnetic configurations in $Li_3Cu_2SbO_6$. The evolution of the X-ESR spectra with temperature is shown in Fig. 3. At high temperature, the spectrum is relatively broad. Considering the full temperature regime, a proper description of the lineshape requires the sum of three Lorentzian functions associated with three resonance modes $L_1$, $L_2$ and $L_3$. Due to the width of the resonance signal, two circular components of the exciting linearly polarized microwave field have to be taken into account for fitting the experimental spectra. Therefore, for the analysis, the ESR signals on both sides of the resonance field $B_r$ have to be included into the fitting formula which was done by applying

$$\frac{dP}{dB} \propto \frac{d}{dB}\left[\frac{\Delta B}{\Delta B^2 + (B-B_r)^2} + \frac{\Delta B}{\Delta B^2 + (B+B_r)^2}\right].$$

(1)

Eq. (1) describes a symmetric line, where P is the power absorbed in the ESR experiment, and $\Delta B$ is the linewidth. The fitted curves (red solid lines in Fig. 3) agree well with the experimental data. A representative spectrum decomposition is shown in the upper panel of Fig. 3 at T = 200 K. The resolved resonance modes are denoted by the colored dashed and dotted lines while their sum is shown by the red solid line.

The temperature dependencies of the effective g-factor, the ESR linewidth and the integral ESR intensity of the three resolved components $L_1$, $L_2$ and $L_3$ as derived from the fitting are collected in Fig. 4. Both the ESR linewidths and the effective g-factors only weakly depend on temperature down to ~ 50 K. The values of the g-factor obtained for $L_1$, $L_2$ and $L_3$ are on average $g_1$ = 2.09 ± 0.02, $g_2$ = 2.11 ± 0.02 and $g_3$ = 2.23 ± 0.04. At T < 50 K, the absorption line broadens and gradually changes into an anisotropic powder pattern, which is typical for a powder sample with magnetic $Cu^{2+}$ ions (see Fig. 3 inset). The corresponding values of the g-tensor are on average $g_\parallel$ = 2.39 ± 0.04 and $g_\perp$ = 2.05 ± 0.04 which is in good agreement with the HF-ESR data (see below).

The difference in the magnetic response for three resolved components in the X-ESR data is clearly seen in the integral intensity of the X-ESR spectra, which is proportional to the number of magnetic spins and was obtained by double integration of absorption derivative curves (Fig. 4c). Both integral ESR intensities $\chi_{L2}$ and $\chi_{L3}$ for the $L_2$ and $L_3$ resonance modes demonstrate Curie-Weiss-like behavior with a small or vanishing Weiss temperature, i.e. $\chi_{L2}$ and $\chi_{L3}$ are associated with weakly interacting or quasi-free spins. In contrast, $\chi_{L1}$ exhibits a broad maximum at T ≈ 100 K resembling the behavior of the spin-gapped contribution to the static magnetic susceptibility $\chi_\Delta$ described above. [32] In contrast, the X-ESR parameters for $L_2$ and $L_3$ increase upon cooling. The similar temperature dependence indicates a common origin of both signals. One may conclude that they reflect the anisotropic powder signal from the quasi-free spin contribution. This assumption is supported by the fact that at lowest temperatures when the signal from spin-gapped subsystem is almost vanished we observe an anisotropic powder-like spectrum.

The HF-ESR data obtained at high frequencies allows to resolve the magnetic anisotropy in more detail. Representative HF-ESR spectra obtained at different frequencies between 80 and 330 GHz, at T = 3.5 K, are shown in Fig. 5. The spectrum at 100 GHz displays an asymmetric but still single resonance feature with a linewidth of ~ 1 T. As frequency increases, the spectrum broadens as expected in an uniaxial magnetic system with two magnetic principal axes corresponding to two components of the g-factor $g_\parallel$ and $g_\perp$. Indeed, the resonance features shown in Fig. 5 are well described by powder simulations applying a common set of effective g-factors, i.e. $g_\parallel = 2.38$ and $g_\perp = 2.05$. The sharp feature in the spectra corresponds to $g_\perp$. Note, that the resonance field of the sharp feature linearly depends on the microwave frequency 80 GHz ≤ f ≤ 330 GHz, with the slope confirming $g_\perp = 2.05$ and without any zero-field splitting (data not shown).

The temperature dependence of the effective g-factors $g_\parallel$ and $g_\perp$ at low temperatures T < 50 K can be clearly observed in the HF-ESR data. As displayed in Fig. 6 which shows the temperature dependence of the HF-ESR spectra at f = 330 GHz, the position of the sharp feature, i.e. $g_\perp$, does not change even down to 4 K. In contrast, broadening of the absorption line below ~50 K indicates increase of $g_\parallel$. This shifting of the effective g-factor $g_\parallel$ shows the evolution of short range spin correlations upon cooling. A similar behavior, i.e. strong changes in $g_\parallel$ but only weak effects on $g_\perp$ are observed, e.g., in the alternating chain compound $Na_2Cu_5Si_4O_{14}$. Another example are the frustrated 1D spin chains in $Li_2CuZrO_4$, where uniaxial anisotropy is associated with similar g-values ($g_\parallel = 2.19$ and $g_\perp = 2.02$), there is only negligible T-dependence of $g_\parallel$ while the evolution of short range order strongly affects $g_\perp$. [33]

It is noted that in the temperature range ~50 K ≤ T ≤ ~100 K where the susceptibility and X-ESR intensity ($L_1$) of the spin-gapped contribution increases to its maximum, the anisotropy of the effective g factor observed in the HF-ESR data does not change significantly. The evolution of the ESR signal with temperature suggests that all Cu ions experience a somehow similar environment of oxygen atoms. The observed values are typical for $Cu^{2+}$ ions which are coordinated by oxygen in a distorted octahedral or approximately square planar manner [34,35].

## 4. Discussion

Our data show the presence of a significant amount of quasi-free magnetic moments as well as of a spin-gapped magnetic subsystem. As mentioned above, $Cu^{2+}$-ions at Li-sites cannot fully account for the Curie-like part of the magnetic susceptibility. Structural refinement however not only implies Cu-ions at the Li-sites but vice versa there are Li-ions replacing Cu-ions in the distorted honeycomb layers. Accordingly, there are about 0.35/f.u. non-magnetic defects in the magnetic layers. We conclude that the Cu-layers with ~17% Li-defects not only cause the spin-gapped contribution to the magnetic susceptibility $\chi_\Delta$ but also account for a part of the Curie-like contribution.

A spin gapped ground state can be an intrinsic property of the honeycomb lattice and a variety of ground states and quantum effects appear upon variation of the relevant magnetic coupling parameters. [36,37] A recent experimental realization is the honeycomb lattice antiferromagnet $InCu_{2/3}V_{1/3}O_3$ which shows anomalous spin dynamics in HF-ESR [38,39]. The g-factor of the trigonal-bipyramidally coordinated $Cu^{2+}$-ions amount to $g_\parallel$ = 2.02 and $g_\perp$ = 2.24 which implies a $3z^2$-$r^2$ ground state in $InCu_{2/3}V_{1/3}O_3$. In contrast, our observation of $g_\parallel$ much larger than $g_\perp$ indicates that the $x^2$-$y^2$ state realized in $Li_3Cu_2SbO_6$ which is typical in plaquette-like cuprates with octahedral symmetry. This confirms the theory results on $Na_3Cu_2SbO_6$ which imply a $x^2$-$y^2$ gound state and associated quasi-1D magnetic exchange interactions as sketched in Fig. 7. [20,21,22,23]

Following these recent works on $Na_3Cu_2SbO_6$, we discuss the static magnetic susceptibility of the spin gapped subsystem $\chi_\Delta$ in terms an alternating S = 1/2 Heisenberg chain, either with alternating AFM exchange interactions or with alternating AFM and FM exchange interactions, $J_1$ and $J_2$ [40].

The magnetic susceptibility hence has been fitted to rational unified expressions

$$\chi(T) = \chi_0 + C/(T+\Theta) + \chi_\Delta(J_1,J_2,N_\Delta)$$

with the temperature-independent term $\chi_0$, the Curie-constant C, the Weiss-temperature $\Theta$, and the number $N_\Delta$ of spins contributing to $\chi_\Delta$. Due to intersite disorder present in the material, the experimental susceptibility data do not enable clearly discriminating the tiny susceptibility differences between AFM-AFM alternating chains and AFM-FM ones with dominant ferromagnetic coupling. However, the microscopic structure suggests ferromagnetic interactions [41] between Cu-ions in adjacent hexagons as the Cu-O-Cu bonding angle amounts to 88.95°. A similar situation with however larger bonding angle of ~94° is realized in the related compound $Na_3Cu_2SbO_6$ where intersite disorder is absent and FM-AFM chains have been identified, e.g., by inelastic neutron scattering, susceptibility studies, and NMR data. [21,22,23].

Fitting the data with a full set of free fitting parameters $\chi_0$, $N_{imp}$, $\Theta$, $J_1$, $J_2$, $N_\Delta$ does not yield reliable results. However, the kink in the inverse susceptibility (Fig. 2) implies $J_1^{AFM}$ = 160 K.

By fixing this value and setting $N_{imp} + N_\Delta = 2$ which is suggested by the synthesis procedure and the XRD analysis, the data are reasonably well described in the AFM-FM chain model [40] by $\chi_0 = 4.6 \cdot 10^{-4}$ erg/(G$^2$mol), $N_{imp} = 0.8$, $\Theta \approx 0$ K, and $|J_2^{FM}| / J_1^{AFM} = 1.78$ (see Fig. 2). We conclude the presence of FM-AFM alternating $S = 1/2$ quantum spin chains in Li$_3$Cu$_2$SbO$_6$ with $J_{FM} = -285$ K and $J_{AFM} = 160$ K. While these results qualitatively agree with recent inelastic neutron data on Na$_3$Cu$_2$SbO$_6$ showing $J_{FM} = -145$ K and $J_{AFM} = 161$ K [21], there is however a striking difference concerning the size of $J_{FM}$. $J_{AFM}$ has a similar value in both systems which agrees to fairly similar Cu-Sb-Cu distances (5.796 Å vs. 5.91 Å). The much larger $J_{FM}$ in Li$_3$Cu$_2$SbO$_6$ can be associated to the Cu-O-Cu bond angle being closer to 90° in Li$_3$Cu$_2$SbO$_6$, i.e. 88.95°, and Cu-Cu distance of 2.92 Å, while these values amount to 95.27° and 2.96 Å in Na$_3$Cu$_2$SbO$_6$ [28]. We also assume, that the large value of $J_{FM}$ is partially due to direct ferromagnetic exchange, similar to what has been found in Li$_2$CuO$_2$ where $J_{FM} = -228$ K. [42]

The resulting FM-AFM spin chain with strongly dominating $J_{FM}$ somehow resembles the scenario of 1D $S = 1$ AFM chains which according to Haldane's conjecture exhibits a spin gap $\Delta$ of about 0.41 J separating the singlet ground state from the triplet excitation spectrum [2,43]. One hence may assign the spin gap found in our susceptibility data with a Haldane-like feature. As mentioned above, this scenario of 1D chains being realized in the hexagonal Cu $S = ½$ layers also involves a significant amount of about 0.35/f.u. non-magnetic defects. Such defects made of Li-ions at Cu-sites will magnetically decouple spins in the 1D chains, essentially resulting in fragmented chains of finite length. In order to estimate the lengths of the chain fragments realized in the honeycomb layers of Li$_3$Cu$_2$SbO$_6$, as a starting point we assume a random distributions of Li-ions at the Cu-sites. According to table S1, the probability to find a Li-atom at any Cu site is $p = 0.17$ while for Cu it is $(1-p) = 0.83$. Hence, the probability to find a sequence of $n$ neighboring Cu-ions is $(1-p)^n$. In order to realize a separated chain fragment, such a sequence must be hedged by two defects, i.e. the probability decreases by $p^2$. Finally, the probability $p_n$ per Li defect to find a $S = ½$ spin chain fragment of length n amount to $p*(1-p)^n$. E.g., the probability to find an isolated $S = ½$ quasi-free spin in the Cu-layer is hence $p_1 = 0.14$ per defect.

The presence of finite Haldane-like chains fragmented by Li defect straightforwardly accounts not only for the gapped spin systems but also for quasi-free moments in the Cu layers. It has been shown for the conventional $S = 1$ Haldane chain system NENP that non-magnetic defects result in two effective S=1/2 spins at both sides of the defects. [44] In general, even (i.e. even number of integer spins) Haldane chain fragments are in the singlet ground state due to effective antiferromagnetic interaction between the edge states while odd spin chains are in the triplet state. It has been however been pointed out by Ohta *et al.*, that, due to finite spin correlations in Haldane chains, S = ½ edge states result in a polarization extending only a few sites from the edge of the chain fragments in $Y_2BaNi_{0.96}Mg_{0.04}O_5$. [12] Hence, while in rather short chains the effective S = 1/2 edge states are forming singlets they may be considered quasi-free if the chain length is long as compared to the correlation length.

Applying this to the case at hand, the Haldane-like scenario can be ignored for the shortest chain fragments with n = 1 and 2. While isolated S = ½ form quasi-free spins, dimers may either form a quasi-free S = 1 situation, if the two S = 1/2 are interacting ferromagnetically, or a spin-gapped antiferromagnetic dimer. Since AFM dimers forming singlets do not matter for the discussion of the S = 1 Haldane-like scenario, the associated probability for relevant, i.e. FM, dimers amounts to $0.5*p_2 \approx 0.06$. In the hypothetical case of infinite AFM correlations lengths, all longer spin chain fragments with n > 2 would be either in the triplet (odd) or singlet state (even). For example, the case of n = 9 includes $N_1 = 4$ ferromagnetic dimers (S = 1) coupled antiferromagnetically. Assuming the correlation length being smaller than 8 average Cu-Cu distances and assuming defects in longer chain fragments resulting in pairs of quasi-free pairs in fragmented Haldane chains as pointed out in Ref. [13], this scenario implies the number of quasi-free S = ½ in the Cu-layers being about 0.35/f.u. [45]. Ferromagnetic dimers in the Cu-layers and Cu-ions on Li sites increase this number to about 0.47/f.u. Together with the quasi-free spins originating from Cu-ions at Li-sites, this value of 0.81/f.u. would perfectly agree to the experimentally observed Curie-like contribution to the magnetic susceptibility. We note that the probability to find $Li^+$ at neighboring sites might be reduced which would yield a slightly larger number than deduced in the simplified model presented here. To summarize, the static magnetic susceptibility data are consistent with the presence of finite FM-AFM chain fragments in the Cu-layers with dominant FM coupling and Haldane-like behavior. The large number of non-magnetic defects results in relative short chain fragments, much shorter than realized in $Y_2BaNi_{0.96}Mg_{0.04}O_5$. [13]

The scenario of finite Haldane-like chains also implies that the spin gap evident in the inset of Fig. 2 would not be *the* spin gap of the differently long chain fragments as the finite size scaling of the singlet-triplet gap amounts to $\Delta = \Delta_0 \cdot \exp(-N_1/\xi)$, with $N_1$ the Haldane chain length ($N_1 = 2n$ here), $\xi$ the correlation lengths, and $\Delta_0 = 1.2\ J_2$ and $2.14\ J_2$ for even and odd chains, respectively. [46] For long chains, which are however rather improbable in the high defect regime (~ 17%) realized in the system studied here, the gap vanishes which reflects the negligible coupling between the $S = ½$ edge states in the case of long chains. We conclude that the observed spin-gapped contribution to the magnetic susceptibility reflects a variety of differently gapped subsystems originating from finite sized Haldane-like chain fragments of different length.

## 4. Conclusions

Layered $Li_3Cu_2SbO_6$ synthesized either by solid-state reaction or by ion exchange is found to exhibit a significant amount of intersite defects. In contrast to previous reports, the system crystallizes in the space group C2/m with halved unit cell volume. Static magnetic susceptibility and ESR measurements show two magnetic contributions, i.e. quasi-free spins at low-temperature and a spin-gapped magnetic subsystem. The data suggest alternating ferromagnetic-antiferromagnetic alternating chains with $J_{FM} = -285$ K and $J_{AFM} = 160$ K. There are about 0.8 quasi-free spins $S = ½$ per formula unit which cannot be explained by $Cu^{2+}$-ions residing on Li positions. Instead, a significant amount of non-magnetic Li-defects in the Cu-chains implies a scenario of highly fragmented Haldane-line chains with the edge-states of the finite Haldane chains contributing the quasi-free moments observed in the experiment.

**Acknowledgments**

The authors are grateful for fruitful discussions with Alexander Tsirlin. Support by the Excellence Initiative of the German Federal Government and States and by the Baden-Württemberg-Stiftung is acknowledged. This work was supported in part from the Ministry of Education and Science of the Russian Federation in the framework of Increase Competitiveness Program of NUST «MISiS» (№ K2-2015-075 and № K4-2015-020) and by Act 211 of the Government of Russian Federation, agreement № 02.A03.21.0006. We acknowledge support from Russian Foundation for Basic Research Grants № 14-02-00245 and 14-03-01122.

**Captions**

Table 1. Comparison of monoclinic (C2/m) lattice parameters and degrees of inversion t (percentage of Li on Cu site) for three different preparations of $Li_3Cu_2SbO_6$

Table 2. Refinement details for two samples of $Li_3Cu_2SbO_6$ (C2/m, Z=2). For lattice parameters see Table 1.

Figure 1. XRD patterns of $Li_3Cu_2SbO_6$ synthesized by solid-state reaction (a) and via ion exchange (b). Symbols shows the experimental data, the black (white) line are simulated data (background), and the red bottom line shows the difference. Vertical ticks correspond to the Bragg positions of the diffraction maxima.

Figure 2. Static magnetic susceptibility $\chi$ = M/B (left ordinate) of $Li_3Cu_2SbO_6$ and its inverse (right ordinate), at B = 0.1 T, vs. temperature. The dashed line represents a Curie-Weiss approximation $\chi_{CW}^I$ at 10 K < T < 30 K to the experimental data $\chi$ and the inset shows the difference $\chi-\chi_{CW}^I$. The dotted line indicates Curie-Weiss-like behavior of experimental data at T > 200 K. Solid (red) line: Approximation by means of the alternating FM-AFM chain model (see the text).

Figure 3. Evolution of the X-ESR spectra with temperature. The symbols are experimental data, and the red solid lines are results of lineshape fitting as described in the text. The upper panel shows a representative ESR spectrum along with the decomposed three resonance modes. The inset presents the spectrum at T = 7 K.

Figure 4. The temperature dependencies of the effective g-factor (a), the ESR linewidth (b) and the integral ESR intensity (c) for the three resolved components $L_1$, $L_2$ and $L_3$ of the X-ESR spectra.

Figure 5. Representative HF-ESR spectra obtained at T = 3.5 K and various frequencies. The red lines represent powder simulations of the spectra with effective g-factors $g_\parallel$ = 2.38 and $g_\perp$ = 2.05.

Figure 6. (a) Temperature dependence of the HF-ESR spectra, at f = 330 GHz. The dashed lines are guide lines showing the evolution of the g-factor with temperature. (b) Effective g-factors $g_\parallel$ and $g_\perp$ vs. temperature as derived from the experimental data by powder simulation of the spectra.

Figure 7: (a) Sketch of the ab-plane of $Li_3Cu_2SbO_6$. Atomic positions are taken from the structural refinement (cf. Fig. 1) while atomic radii are not to scale and Li-ions acting as non-magnetic defects are distributed by hand in order to illustrate intersite disorder. Dashed and solid lines indicate the dominating magnetic exchange interactions. The grey hexagon illustrates the underlying honeycomb structure. (b) Schematic of the minimal magnetic model.

**Tables**

Table 1. Comparison of monoclinic (C2/m) lattice parameters and degrees of inversion t (percentage of Li on Cu site) for three different preparations of $Li_3Cu_2SbO_6$

| Source | a, Å | b, Å | c, Å | β, ° | V, Å$^3$ | t, % |
|---|---|---|---|---|---|---|
| SSS [25] reindexed as C2/m [24] | 5.4655(7) | 8.7216(8) | 5.3845(7) | 115.270(5) | 232.1 | 20 |
| This work, SSS | 5.4713(1) | 8.7210(1) | 5.3870(1) | 115.315(1) | 232.36(1) | 16.7(1) |
| This work, IEP | 5.4824(3) | 8.7214(3) | 5.3921(2) | 115.340(3) | 233.01(2) | 15.7(2) |

Table 2. Refinement details for two samples of $Li_3Cu_2SbO_6$ (C2/m, Z=2). For lattice parameters see Table 1.

|  | SSS | IEP |
|---|---|---|
| 2Θ range, ° | 15-105.6 | 15-90 |
| number of points | 4531 | 3751 |
| number of hkl | 157 | 102 |
| number of parameters | 60 | 60 |
| R | 0.0385 | 0.0392 |
| wR | 0.0494 | 0.0497 |
| wR expected | 0.0421 | 0.0471 |
| R(F$^2$) | 0.04036 | 0.02277 |
| $\chi^2$ | 1.393 | 1.132 |

**Figures**

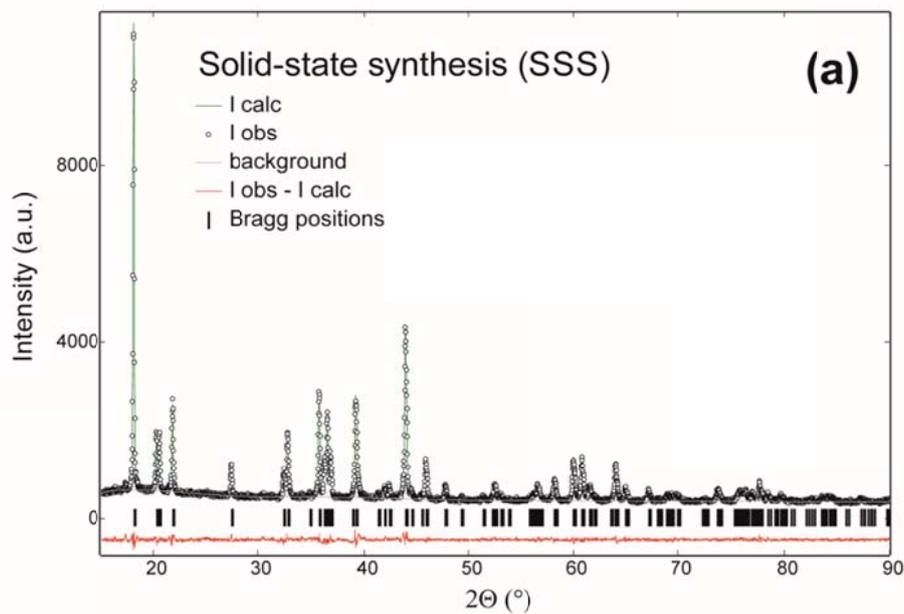

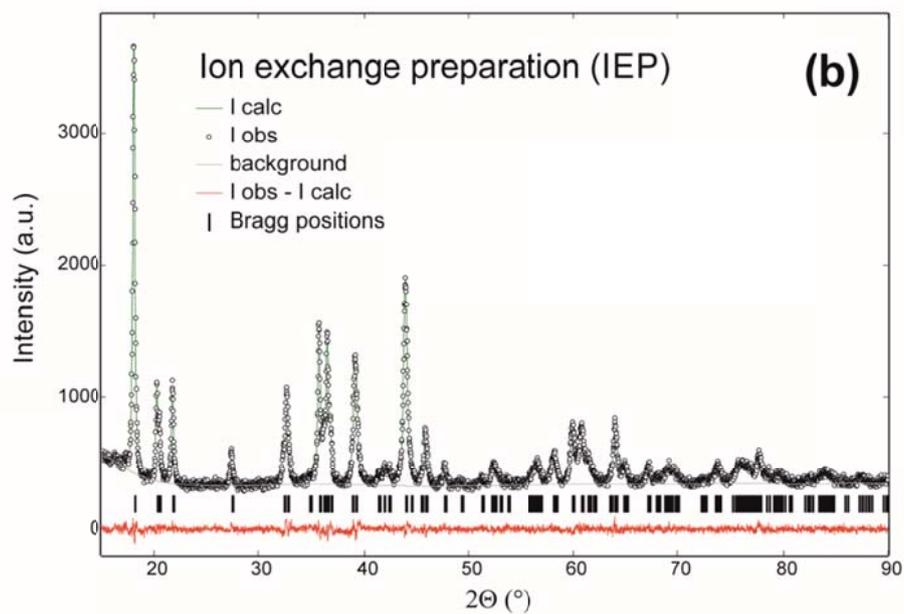

Figgure 1. XRD patterns of $Li_3Cu_2SbO_6$ synthesized by solid-state reaction (a) and via ion exchange (b). Symbols shows the experimental data, the black (white) line are simulated data (background), and the red bottom line shows the difference. Vertical ticks correspond to the Bragg positions of the diffraction maxima.

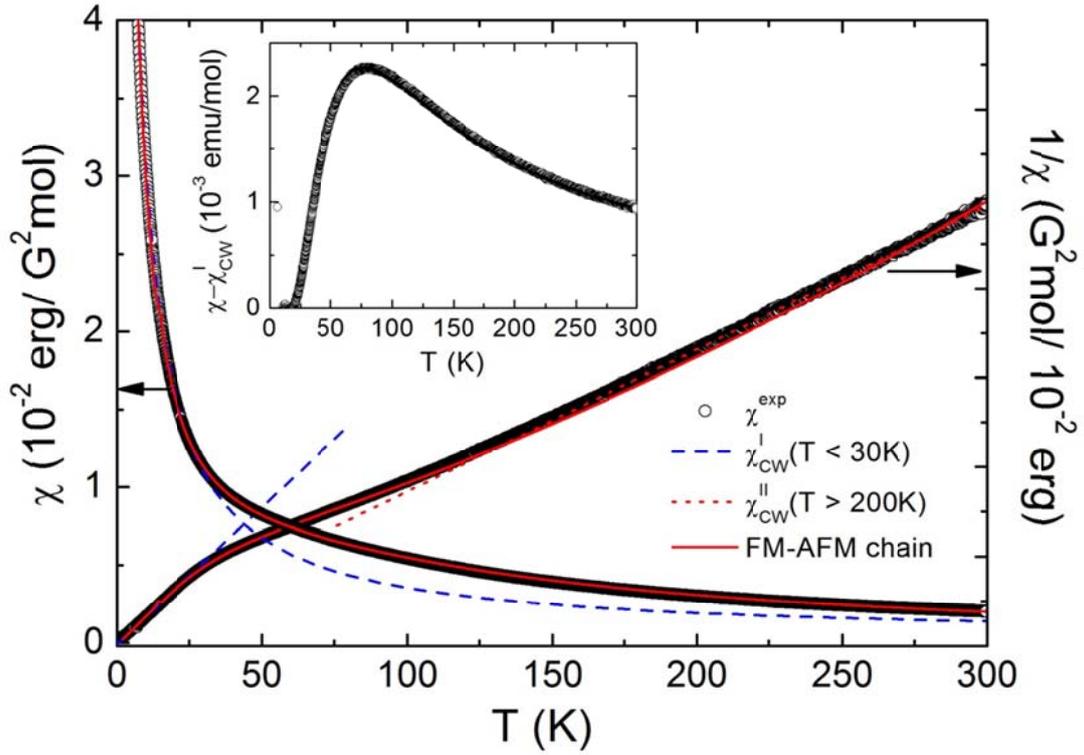

Figure 2. Static magnetic susceptibility $\chi = M/B$ (left ordinate) of $Li_3Cu_2SbO_6$ and its inverse (right ordinate), at $B = 0.1$ T, vs. temperature. The dashed line represents a Curie-Weiss approximation $\chi_{CW}^I$ at 10 K < T < 30 K to the experimental data $\chi$ and the inset shows the difference $\chi - \chi_{CW}^I$. The dotted line indicates Curie-Weiss-like behavior of experimental data at T > 200 K. Solid (red) line: Approximation by means of the alternating FM-AFM chain model (see the text).

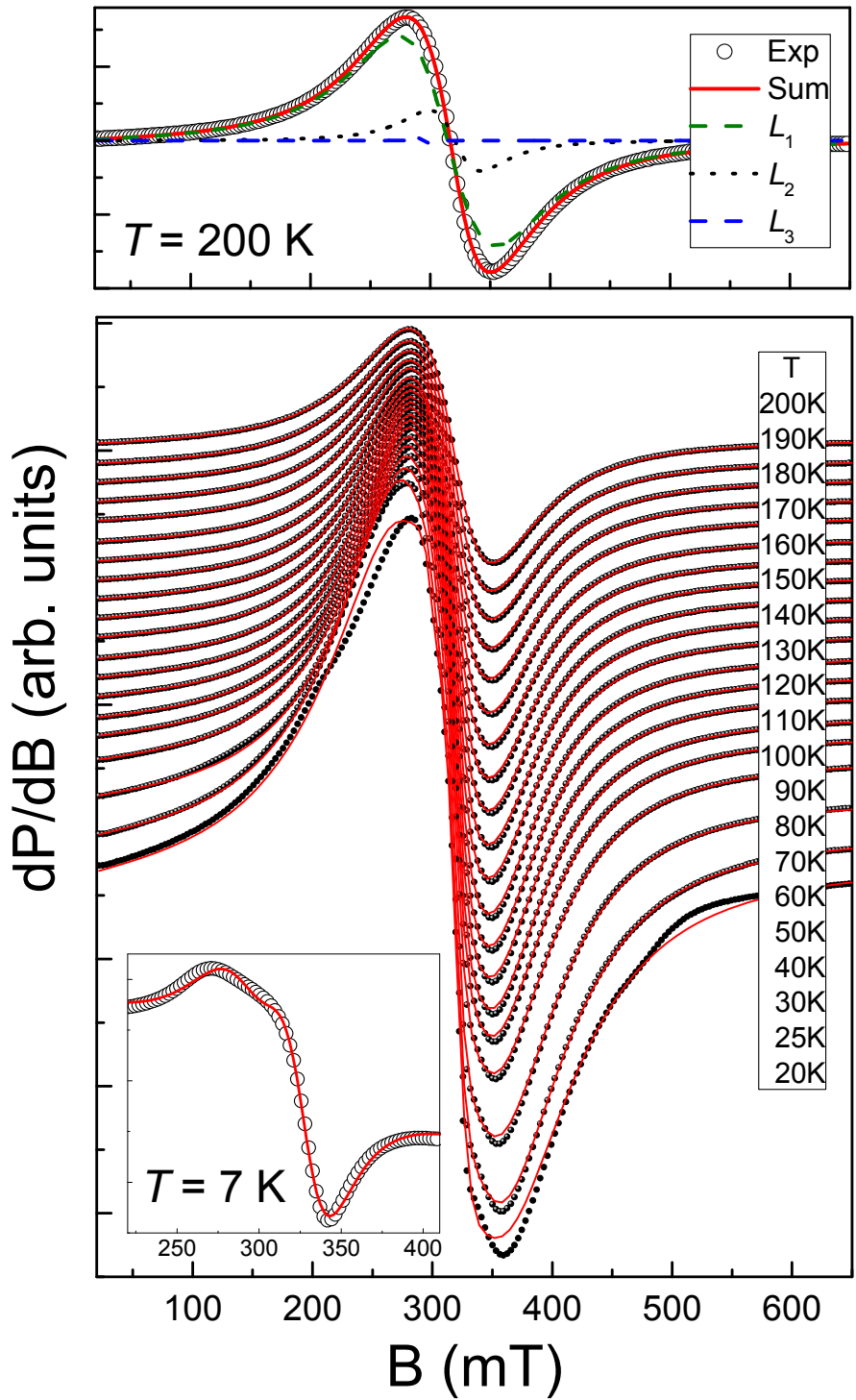

Figure 3. Evolution of the X-ESR spectra with temperature. The symbols are experimental data, and the red solid lines are results of lineshape fitting as described in the text. The upper panel shows a representative ESR spectrum along with the decomposed three resonance modes. The inset presents the spectrum at T = 7 K.

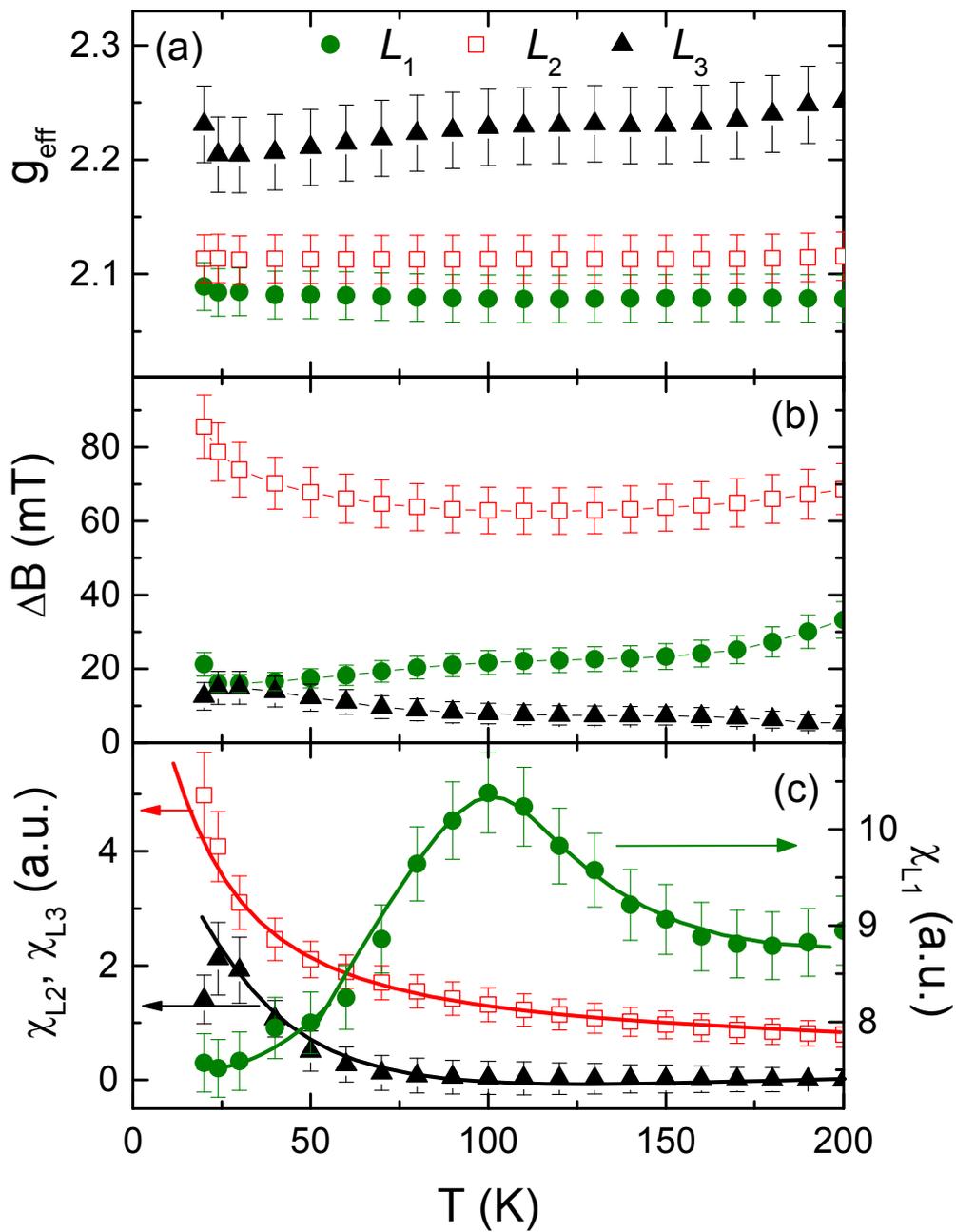

Figure 4. The temperature dependencies of the effective g-factor (a), the ESR linewidth (b) and the integral ESR intensity (c) for the three resolved components $L_1$, $L_2$ and $L_3$ of the X-ESR spectra. The lines are guides to the eye.

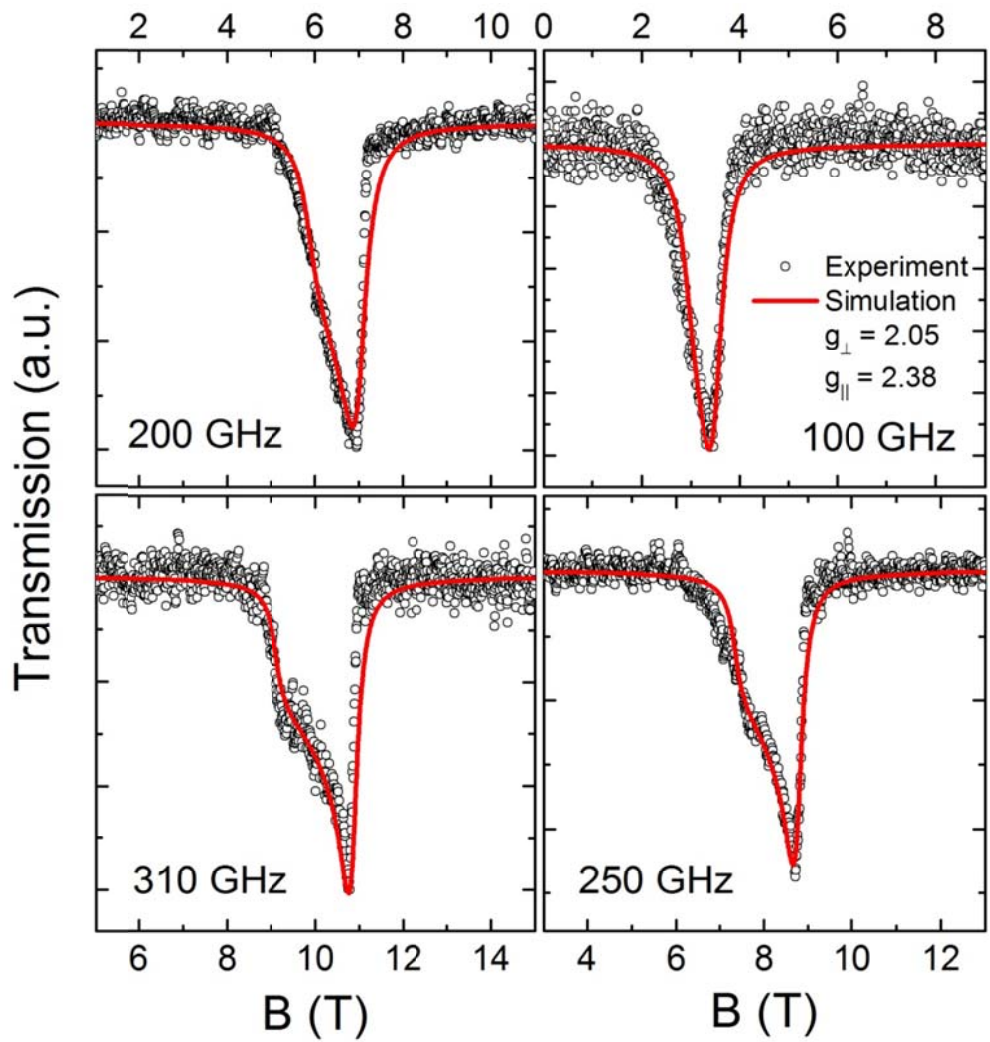

Figure 5. Representative HF-ESR spectra obtained at T = 3.5 K and various frequencies. The red lines represent powder simulations of the spectra with effective g-factors $g_\parallel$ = 2.38 and $g_\perp$ = 2.05.

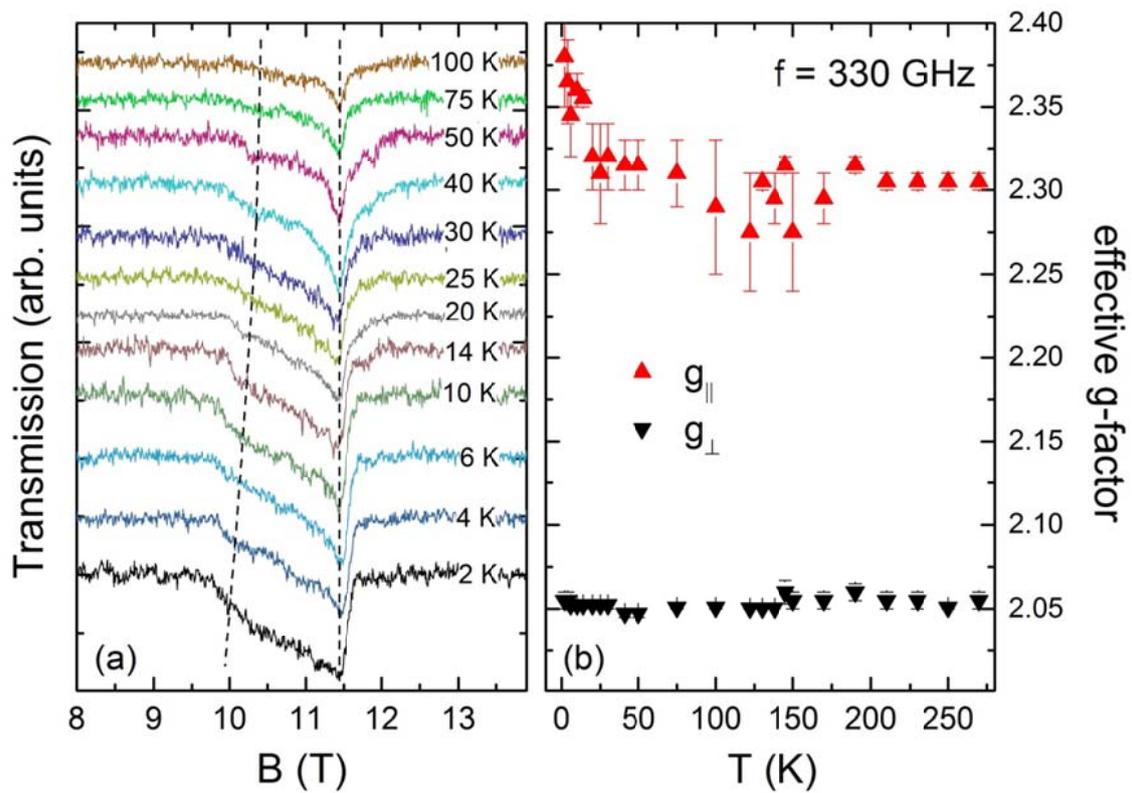

Figure 6. (a) Temperature dependence of the HF-ESR spectra, at f = 330 GHz. The dashed lines are guide lines showing the evolution of the g-factor with temperature. (b) Effective g-factors $g_\parallel$ and $g_\perp$ vs. temperature as derived from the experimental data by powder simulation of the spectra.

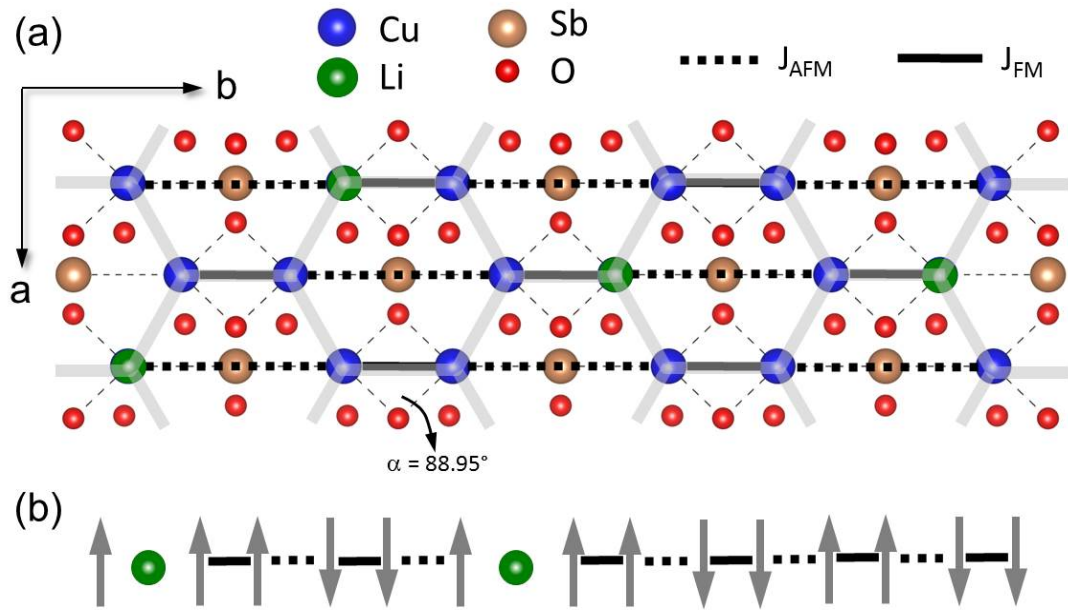

Figure 7: (a) Sketch of the ab-plane of $Li_3Cu_2SbO_6$. Atomic positions are taken from the structural refinement (cf. Fig. 1) while atomic radii are not to scale and Li-ions acting as non-magnetic defects are distributed by hand in order to illustrate intersite disorder. Dashed and solid lines indicate the dominating magnetic exchange interactions. The grey lines illustrate the underlying honeycomb structure. (b) Schematic of the minimal magnetic model.